\documentclass[twocolumn,amsmath,amssymb,pra]{revtex4-1}

\usepackage{graphicx}
\usepackage{color}
\usepackage{dcolumn}
\usepackage{amsmath}
\usepackage{subfigure}
\usepackage{sidecap}

\usepackage{mathrsfs}
\usepackage{amsfonts}
\usepackage{indentfirst}
\usepackage{bm}
\usepackage{dsfont}

\usepackage[colorlinks,citecolor=blue]{hyperref}

\newcommand{\be}{\begin{equation}}
\newcommand{\ee}{\end{equation}}
\newcommand{\bey}{\begin{eqnarray}}
\newcommand{\eey}{\end{eqnarray}}
\newcommand{\bw}{\begin{widetext}}
\newcommand{\ew}{\end{widetext}}

\newcommand{\ra}{\rangle}
\newcommand{\la}{\langle}

\newcommand{\ba}{\begin{array}}
\newcommand{\ea}{\end{array}}
\newcommand{\bi}{\begin{itemize}}
\newcommand{\ei}{\end{itemize}}
\newcommand{\bem}{\begin{enumerate}}
\newcommand{\eem}{\end{enumerate}}

\begin{document}

\title{Excited state quantum phase transition and Loschmidt echo spectra 
in a spinor Bose-Einstein condensate}

\author{Zhen-Xia Niu$^{1}$}
\author{Qian Wang$^{1,2}$}\email{qwang@zjnu.edu.cn}

\affiliation{$^{1}$Department of Physics, Zhejiang Normal University, Jinhua 321004, China \\
$^2$CAMTP-Center for Applied Mathematics and Theoretical Physics, University of Maribor,
Mladinska 3, SI-2000, Maribor, Slovenia}

\date{\today}

\begin{abstract}

Identifying dynamical signatures of excited state quantum 
phase transitions (ESQPTs) in experimentally 
realizable quantum many-body systems is helpful for 
understanding the dynamical effects of ESQPTs.
In such systems, the highly controllable spinor Bose-Einstein condensates (BECs) 
offer an exceptional platform to study ESQPTs. 
In this work, we investigate the dynamical characteristics of the ESQPT 
in spin-$1$ BEC by means of the Loschmidt echo spectrum. 
The Loschmidt echo spectrum is an extension of the 
well-known Loschmidt echo and defined as the overlaps between 
the evolved state and the excited states of the initial Hamiltonian.
We show that both the time evolved and long time averaged Loschmidt echo spectrum
undergo a remarkable change as the system passes through the critical point of the ESQPT.
Moreover, the particular behavior exhibited by the Loschmidt echo spectrum at the critical point
stands as a dynamical detector for probing the ESQPT. 
We further demonstrate how to capture the features of the ESQPT by using the energy 
distribution associated with the Loschmidt echo spectrum for time evolved 
and long time averaged cases, respectively. 
 Our findings contribute to a further verification of the usefulness of the Loschmidt echo spectrum
 for witnessing various quantum phase transitions in many-body systems and
 provide a different way to experimentally examine the dynamical consequences of ESQPTs.

\end{abstract}

\maketitle

 \section{Introduction}

 Understanding the notion of excited state quantum phase transitions (ESQPTs) 
 in quantum  many-body systems has attracted a lot of interest in recent years 
 \cite{Cejnar2006,Caprio2008,Stransky2014,Stransky2016,Cejnar2021,Corps2021}.
 As a generalization of the ground state quantum phase transition (QPT) 
 \cite{Sachdev2011} to excited states, 
 ESQPTs exist in a wide range of quantum many-body systems, 
 including the interacting boson model \cite{Caprio2008,PerezB2008,Macek2019,DongW2021},
 molecular bending transitions \cite{Larese2013,Jamil2021,Jamil2022}, 
 the kicked and coupled top models \cite{Bastidas2014,Garcia2021,WangP2021,Mondal2022},
 the Kerr-nonlinear oscillator \cite{WangS2020,Carlos2022},
 the spinor Bose-Einstein condensates (BECs)
 \cite{TianT2020,Feldmann2021,Cabedo2021,Meyer2023,ZhouL2022},
 the Lipkin-Meshkov-Glick (LMG) model
 \cite{Caprio2008,Leyvraz2007,Sindelka2017,Nader2021,Gamito2022,Santos2015,Santos2016}, 
 as well as the Dicke and Rabi models 
 \cite{PerezF2011,Brandes2013,Puebla2013,Magnani2014,Puebla2016}.
 
 ESQPTs are characterized by singularities in the density of states (DOS).
 The energy that leads to the divergence of the DOS is identified as the ESQPT critical energy.
 In contrast to the standard QPT, 
 which describes an unsmooth variation
 of a system's ground state properties with a control parameter,
 an ESQPT shows strong effects on a large amount 
 of excited states and results in non-analytical evolution in their structure and energy 
 with both the energy and control parameter. 
 Although ESQPTs extend the QPTs to excited states, 
 they can take place in the systems with no QPT \cite{Relano2016,Stransky2021,Corps2022}.  
 The ESQPT occurs only in the thermodynamic limit, but its emergence can be revealed by scaling
 analyses in finite systems.
 It is worth pointing out that various studies on ESQPTs are focused on the systems 
 that are associated with 
 classical counterparts having a few degrees of freedom, however, the definition of ESQPTs for 
 any number of degrees of freedom was investigated by the authors of Ref.~\cite{Stransky2016}. 
 
 The close connection between ESQPTs and several fundamental questions in diverse areas of modern
 physics has led the study of ESQPTs to become a quite active research field.  
 There is a vast amount of theoretical studies on the effects of ESQPTs.
 It was known that ESQPTs can strongly affect the quantum dynamics 
 after a quench
 \cite{PerezF2011,WangP2021a,Puebla2015,WangP2019a,
 WangP2019b,Relano2008,Relano2009,Stransky2021,Kloc2021}, 
 localize the eigenstates \cite{Santos2015,Santos2016}, accelerate the time evolution of the system 
 \cite{Kloc2018,Hummel2019,Cameo2020},
 and create the Schr\"{o}dinger cat states \cite{Corps2022}.
 ESQPTs are also closely related to the onset of chaos 
 \cite{Garcia2021,Fernandez2011b,Lobez2016,Corps2022a} 
 and different types of dynamical quantum phase transitions
 \cite{Corps2022b,Corps2022c,Corps2022d}, 
 as well as the thermal phase transition \cite{PerezF2017}. 
 The particular role played by the ESQPTs in the thermalization process of isolated quantum systems
 has also been revealed recently \cite{Kelly2020,Lambert2022}.
 Moreover, the endeavour to explore the order parameters of 
 ESQPTs \cite{Corps2021,Puebla2013,WangP2019b,Feldmann2021,PueblaA2013} 
 provides further insights into the excited state phases.
 
 To date, ESQPTs were experimentally observed in
 molecular bending transitions \cite{Larese2013, Jamil2021}
 and microwave Dirac billiards \cite{Dietz2013} via the singularities in the DOS.
 However, none of these systems allows us to experimentally probe
 the dynamical signatures of ESQPTs in a controllable way.  
 In this context, spinor BECs \cite{Kawaguchi2012,SKurn2013} provide excellent platforms 
 to study the dynamical signatures of ESQPTs, because
 they have a high degree of controllability
 \cite{Stenger1998,Sadler2006,Gerbier2006,Hamley2012,ZhaoL2014,XYLuo2017} 
 and show ESQPTs in their spectra
 \cite{TianT2020,Feldmann2021,Cabedo2021,ZhouL2022}. 
 While signatures of the ground state QPT and associated dynamical features
 in spinor BECs were theoretically analyzed
 \cite{Kawaguchi2012,SKurn2013,Dag2018,WenxianZ2003,ZhangZ2013,XueM2018} 
 and observed in numerous experiments
 \cite{XYLuo2017,Jacob2012,Anquez2016,LiuY2009,Bookjans2011,JianJ2014,Vinit2017,HYLiang2021},
 only a few recent studies were concerned with the 
 characteristics of ESQPTs \cite{Feldmann2021,Cabedo2021,Meyer2023}.

 In this work, we investigate the dynamical signatures of ESQPT in the spin-$1$ BEC using the 
 Loschmidt echo spectrum, which was proposed in a very recent work \cite{Wong2022}.
 The Loschmidt echo spectrum extends the conventional concept of the Loschmidt echo beyond the
 ground state and is defined as the overlaps between the evolved state and the excited states.
 As the dynamics in a quantum system is governed by its entire energy spectrum, 
 one can expect that the Loschmidt echo spectrum will offer more 
 dynamical characteristics of the system.
 Since the ESQPT typically affects the full spectrum of excited states, 
 it is expected that more insights into the dynamical properties of ESQPT 
 can be obtained through the Loschmidt echo spectrum. 
 The usefulness of the Loschmidt echo spectrum in studying of
 the dynamical quantum phase transitions (DQPTs) in spin systems 
 has been verified \cite{Wong2022}, which shows that the Loschmidt echo spectrum
 not only behaves as a powerful detector of DQPTs 
 but also helps us to get a better understanding on the physical nature of DQPTs. 
 Here, we examine its ability to characterize the ESQPT.
 To the best of our knowledge, the Loschmidt echo spectrum has not been used to
 explore the dynamical signatures of the ESQPT in previous works.
 
 The ESQPT in the spin-$1$ BEC is signified by the logarithmic divergence of the DOS
 and it can lead to an abrupt change in the structure of eigenstates.
 The analysis of the classical limit of the system shows that the onset of ESQPT can be considered 
 as a consequence of the saddle point in its classical counterpart. 
 We show that the underlying ESQPT exhibits strong impact on the time evolution of the Loschmidt
 echo spectrum. 
 Both the existence and different phases of the ESQPT can be reliably identified by the properties of the
 Loschmidt echo spectrum and associated energy distribution.
 We further display how the ESQPT manifests itself in the long time averaged Loschmidt 
 echo spectrum and associated energy distribution. 
 The aim of the present work is to access the signatures of ESQPT 
 from both time evolved and long time averaged behaviors of the Loschmidt echo spectrum, 
 as well as to provide further evidence
 of the validity of the Loschmidt echo spectrum for diagnosing various phase transitions in 
 quantum many-body systems. 
 
 The rest of the article is structured as follows.
 In Sec.~\ref{secondS}, we introduce the considered physical system 
 and its classical limit and discuss the basic features of the ESQPT. 
 Our main results are presented in Sec.~\ref{thirdS}, where 
 we report the properties of the Loschmidt echo spectrum 
 and associated energy distribution, and reveal their connection with the ESQPT 
 for time evolved and long time averaged cases, respectively.  
 Finally, we summarize our results and conclude in Sec.~\ref{summary}.

  \begin{figure}
  \includegraphics[width=\columnwidth]{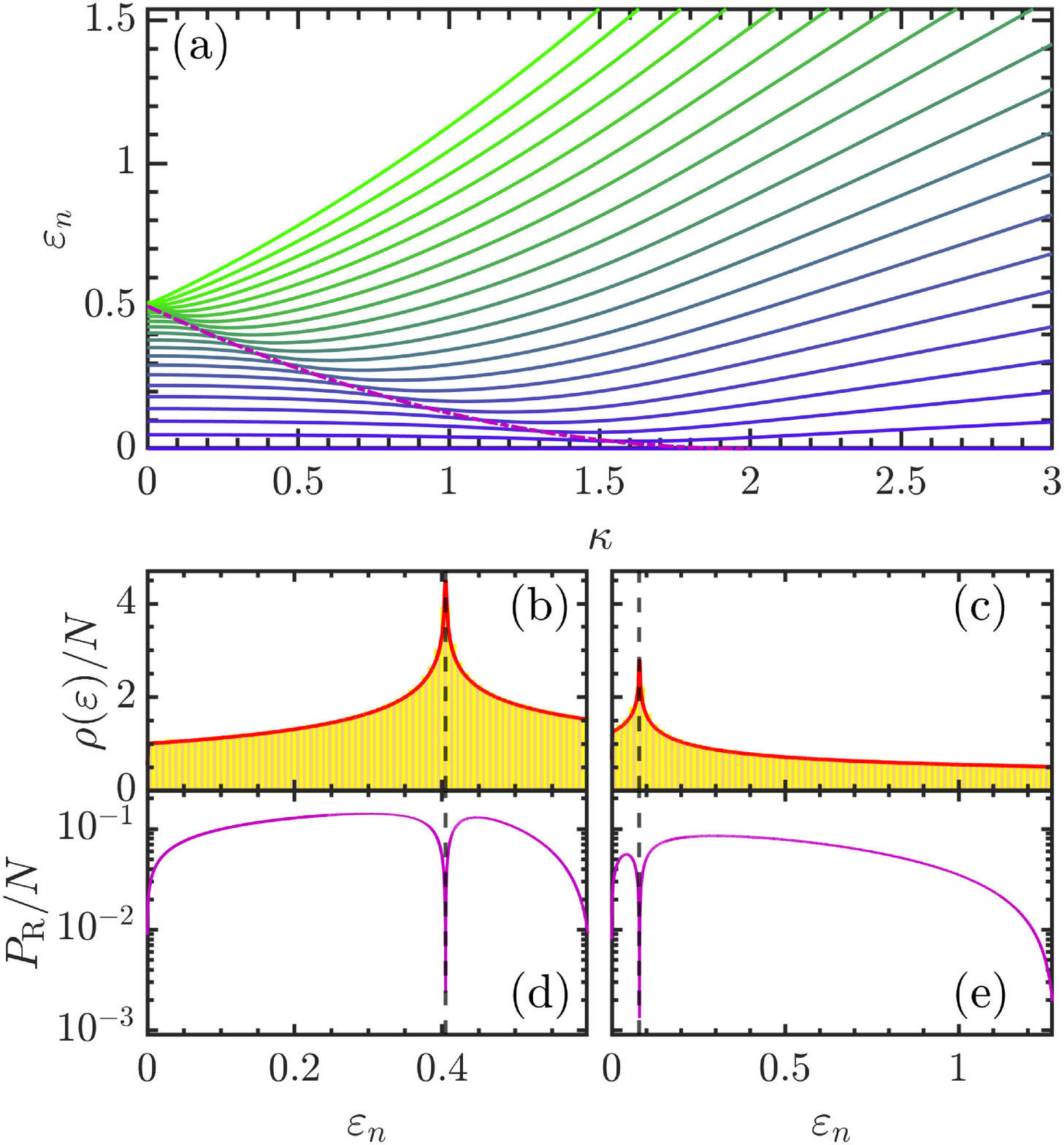}
  \caption{(a): Rescaled energy levels $\varepsilon_n=(E_n-E_0)/N$ as a function of $\kappa$ 
  with $N=40$. The purple dot-dashed line denotes the critical energy of ESQPT, 
  as given in Eq.~(\ref{CrtE}).
  (b)-(c): Normalized density of states $\rho(\varepsilon)/N$ for $\kappa=0.2$ (b) 
  and $\kappa=1.2$ (c) with $N=10000$.
  The red solid lines are the analytical results, see Eq.~(\ref{DOSHc}).
   (d)-(e): Rescaled participation ratio $P_\mathrm{R}/N$ 
   as a function of $\varepsilon_n$ for the same values of $\kappa$ 
   as in panels (b) and (c) with $N=10000$.
  The vertical dashed lines in panels (b)-(e) represent the critical energy of the ESQPT obtained from
  Eq.~(\ref{CrtE}).}
  \label{SpDos}
 \end{figure}

 \section{Quantum system}  \label{secondS}
 
 We consider the spin-$1$ BEC, which describes $N$ mutually interacting atoms 
 with three hyperfine spin states $m=0,\pm1$ 
 and can be experimentally realized by 
 $^{87}\mathrm{Rb}$ and $^{23}\mathrm{Na}$ atoms \cite{Kawaguchi2012,SKurn2013}.
 Using the single-mode approximation \cite{Kawaguchi2012} 
 which decouples the spin and spatial degrees
 of freedom, the Hamiltonian of the system can be written as \cite{Kawaguchi2012,SKurn2013}
 \begin{align}\label{spinorH}
    \frac{\hat{H}}{|c|}&=\frac{\mathrm{sign}(c)}{N}\left[\hat{a}_0^{\dag2}\hat{a}_{1}\hat{a}_{-1}
    +\hat{a}_1^\dag\hat{a}_{-1}^\dag\hat{a}_0^2
      +\hat{N}_0(\hat{N}_1+\hat{N}_{-1})\right. \notag \\
    &\left.+\frac{1}{2}(\hat{N}_1-\hat{N}_{-1})^2\right]+\kappa(\hat{N}_1+\hat{N}_{-1}).
 \end{align}
 Here, $\hat{a}_m (\hat{a}_m^\dag)$ with $m=0,\pm1$ are the bosonic annihilation (creation)
 operators for spin state $m$, $\hat{N}_m=\hat{a}^\dag_m\hat{a}_m$ denote
 the atom number operators and satisfy $\sum_m\hat{N}_m=\hat{N}$.
 The parameter $c$ refers to the strength of the inter-spin interaction 
 with $c<0\ (c>0)$ for ferromagnetic (antiferromagnetic) BEC
 \cite{Kawaguchi2012}, while $\kappa\equiv q/|c|$ represents the 
 effective quadratic Zeeman shift and can 
 take both positive and negative values by means of microwave dressing \cite{Gerbier2006,ZhaoL2014}.

 The Hamiltonian (\ref{spinorH}) conserves the magnetization 
 $\hat{\mathcal{M}}_z=\hat{N}_1-\hat{N}_{-1}$ and parity $(-1)^{\hat{N}_0}$ \cite{Feldmann2021}. 
 In this work, we restrict our analysis to the even parity subspace with $\mathcal{M}_z=0$, thus
 the Hilbert space has dimension $\mathcal{D}_\mathcal{H}=N/2+1$.
 Moreover, we only consider the ferromagnetic spinor BEC with $c<0$ and $\kappa\geq0$.
 However, we would like to point out that it is straightforward to generalize our studies to the cases
 of antiferromagnetic spinor condensate, $\mathcal{M}_z\neq0$ and $\kappa<0$. 
 
 For our considered case,
 it is known that the system exhibits a ground state QPT at 
 the critical point $\kappa_c=2$, which divides the broken-axisymmetry phase 
 with $\kappa<2$ from the polar phase with $\kappa>2$
 \cite{SKurn2013,WenxianZ2003,Jacob2012,ZhangZ2013,Anquez2016,XueM2018}. 
 In addition, the existence of ESQPTs and the relevant signatures and applications 
 in spinor BECs were also investigated in recent works 
 \cite{Feldmann2021,TianT2020,Cabedo2021,ZhouL2022}.
 
 The main characteristics of ESQPT in spin-1 BEC have been unveiled \cite{Feldmann2021}.
 In the following subsections, we will briefly review the results 
 in Ref.~\cite{Feldmann2021} and perform further analysis on the features of the ESQPT 
 by focusing on the eigenstates' localization property in both the quantum and classical cases.

\subsection{Excited state quantum phase transitions} \label{SecSb}

ESQPTs are signified by the clustering of the eigenlevels, which results in 
the singularities in the DOS \cite{Caprio2008,Cejnar2021}.
To illustrate the ESQPT in the spin-1 BEC, we analyze the spectrum of Hamiltonian (\ref{spinorH}) 
as a function of control parameter $\kappa$.
In Fig.~\ref{SpDos}(a), we show how the rescaled excitation energies, $\varepsilon_n=(E_n-E_0)/N$,
evolve with increasing $\kappa$. Here, $E_n$ denotes the $n$th 
eigenenergy of $\hat{H}$ and $E_0$ is its ground state energy.
One can see that, the excited eigenlevels exhibit an obvious cluster along the dot-dashed line, which
marks the critical energy of the ESQPT, denoted by $\varepsilon_c$. 
We further observe that the critical energy $\varepsilon_c$ decreases with increasing $\kappa$ and 
moves towards to the known ground state QPT as $\kappa$ approaches $2$. 
By utilizing the mean-field approximation, the explicit dependence of the 
critical energy $\varepsilon_c$ on the control parameter $\kappa$ 
can be obtained analytically [see Eq.~(\ref{CrtE}) below].
As a consequence of the eigenlevels cluster, the DOS,
$\rho(\varepsilon)=\sum_n\delta(\varepsilon-\varepsilon_n)$, 
shows a sharp peak at $\varepsilon_c$ for the cases
of $\kappa<2$, as visualized in Figs.~\ref{SpDos}(b) and \ref{SpDos}(c).
In the $N\to\infty$ limit, the peak in the DOS turns into a logarithmic divergence, which makes up
a prominent signature of the ESQPT.

The emergence of the ESQPT is also closely connected with the localization of the eigenstates,
namely, the eigenstates in the neighborhood of $\varepsilon_c$
 are the highly localized states \cite{Santos2015,Santos2016}. 
 The degree of localization of the eigenstates $|\psi_n\rangle$ for a given basis $\{|\alpha_k\rangle\}$,
 is measured by the participation ratio
 $P_\mathrm{R}=1/\sum_k|c_k^{(n)}|^4$, where $c_k^{(n)}=\langle \alpha_k|\psi_n\rangle$. 
 For the spin-1 BEC, the basis consists of the Fock states $|\alpha_k\rangle=|N_{-1},N_0,N_1\rangle$.
 The extended state has a larger value of $P_\mathrm{R}$, while a small $P_\mathrm{R}$ indicates
 the localized state.
 Figs.~\ref{SpDos}(d) and \ref{SpDos}(e) plot the rescaled participation ratio as a function of 
 $\varepsilon_n$ for different control parameters.
 A sharp dip in the behavior of $P_\mathrm{R}$ at $\varepsilon_n\sim\varepsilon_c$ is clearly visible.
 The appearance of dip in $P_\mathrm{R}$ means that the eigenstate at the critical 
 energy is localized in the Fock basis.

\subsection{Classical limit}

More insights into the characteristics of the ESQPT can be obtained by analyzing the properties of the 
stationary points in the corresponding classical system.
To derive the classical counterpart of the Hamiltonian (\ref{spinorH}), we first note that
in the classical limit with $N\to\infty$, the spin states are described by the coherent states
\cite{Feldmann2021,ZhouL2022,Rautenberg2020}
\be
   |\bm{\xi}\rangle=\frac{1}{\sqrt{N!}}
   \left(\xi_{-1}a_{-1}^\dag+\xi_0a_0^\dag+\xi_1a_1^\dag\right)^N|0\ra,
\ee
where $|0\ra$ is the vacuum state, $\xi_m=\sqrt{N_m}e^{i\varphi_m}$ 
with $\varphi_m\in[-\pi,\pi)$ and $\sum_mN_m=N$.
 Then by using the relation $\la\bm{\xi}|a_m|\bm{\xi}\ra=\xi_m$, it is straightforward to find that 
the classical Hamiltonian with $N_{-1}=N_1$ is given by \cite{Kawaguchi2012,Feldmann2021,ZhangW2005}
\be \label{ClassicalH}
   H_{cl}=\lim_{N\to\infty}\frac{\la\bm{\xi}|\hat{H}|\bm{\xi}\ra}{N}=
      \kappa(1-\zeta_0)-2\zeta_0(1-\zeta_0)\cos^2\varphi,
\ee 
where $\zeta_0=N_0/N$ and $\varphi=\varphi_0-(\varphi_1+\varphi_{-1})/2$.
The classical equations of motion are, therefore, given by
\begin{align}
  &\dot{\varphi}=\frac{\partial H_{cl}}{\partial{\zeta_0}}=-\kappa-2(1-2\zeta_0)\cos^2\varphi, 
  \notag \\
  &\dot{\zeta}_0=-\frac{\partial H_{cl}}{\partial{\varphi}}=-2\zeta_0(1-\zeta_0)\sin(2\varphi),
\end{align}
with the constrained condition $d(\varphi_{1}-\varphi_{-1})/dt=0$.

   \begin{figure}
   \includegraphics[width=\columnwidth]{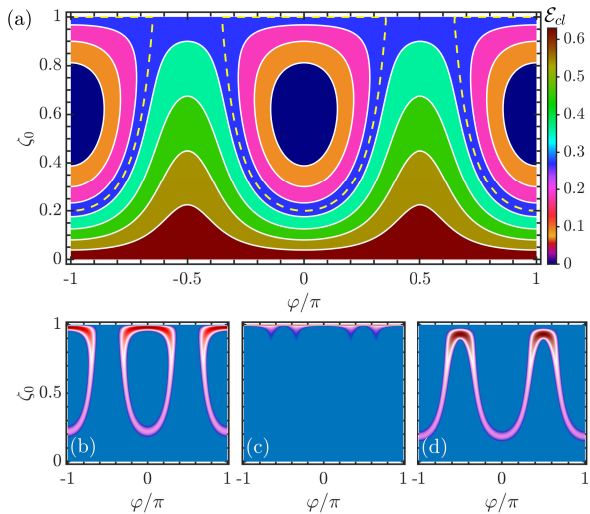}
   \caption{(a): Energy contours of the classical Hamiltonian (\ref{ClassicalH}) in the phase space.
   The yellow dashed line denotes the separatrix given by Eq.~(\ref{CrtE}).
   (b)-(d): Rescaled Husimi distributions, $\mathcal{Q}(\zeta_0,\varphi)=
   Q(\zeta_0,\varphi)/Q_{max}(\zeta_0,\varphi)$ for $\varepsilon_n=0.2909$ (b), 
   $\varepsilon_n=\varepsilon_c=0.32$ (c), and 
   $\varepsilon_n=0.347$ (d) with $N=300$.
   Other parameter: $\kappa=0.4$.}
   \label{Hsprb}
 \end{figure}

By setting $\dot\varphi=\dot{\zeta}_0=0$, one can find that the classical system (\ref{ClassicalH}) has
three fixed points when $\kappa<2$.
They are two stable points $\{\cos\varphi,\zeta_0\}=\{\pm1, (2+\kappa)/4\}$ with the 
minimal energy of the classical system $\mathcal{E}_{min}=-(\kappa-2)^2/8$,
and a saddle point at $\zeta_0=1$ with energy $\mathcal{E}_s=0$.
In Fig.~\ref{Hsprb}(a), we plot the energy contours in phase space for the classical system.
We observe two different structures in the energy surface.  
The change in the structure of the classical energy surface indicates the presence of the separatrix in the
classical dynamics.
The equation of the separatrix in Fig.~\ref{Hsprb}(a) is determined by the energy difference 
$\mathcal{E}_s-\mathcal{E}_{min}$, which also corresponds to the critical energy of the ESQPT,
\be\label{CrtE}
    \varepsilon_c=\frac{(\kappa-2)^2}{8},
\ee
with $0<\kappa<2$.
This result was plotted as the purple dot-dashed line in Fig.~\ref{SpDos}(a).

  \begin{figure*}
   \includegraphics[width=\textwidth]{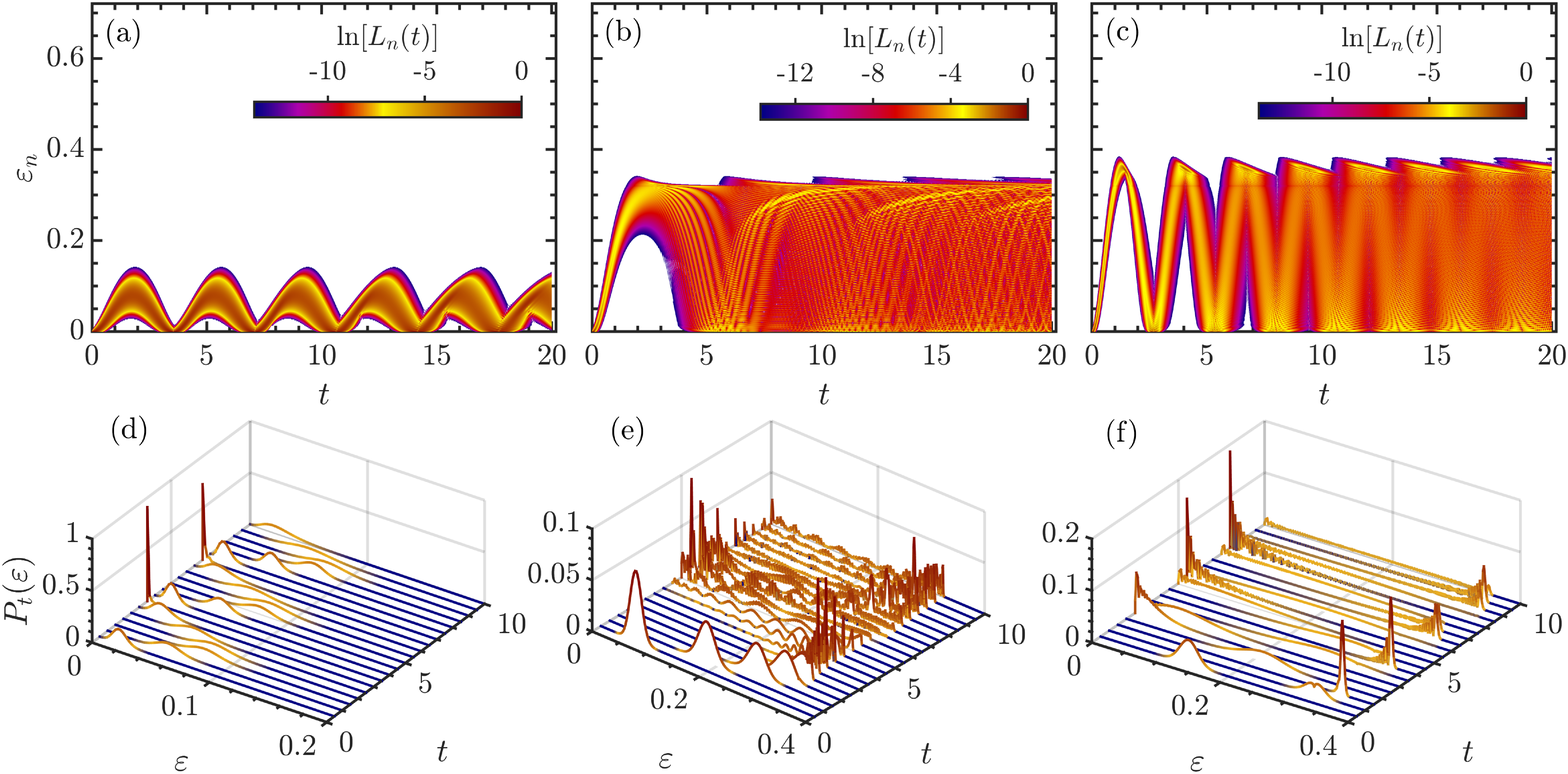}
   \caption{(a)-(c) Color plot of the Loschmidt echo spectrum (in logarithmic scale) 
   as a function of $t$ and $\varepsilon_n$ for 
   $\delta\kappa=0.4$ (a), $\delta\kappa=\delta\kappa_c=0.8$ (b), and $\delta\kappa=1.2$ (c).
   In each panel, the white regions denote $L_n(t)=0$.
   (d)-(f) Evolving of the energy distribution  
   $P_t(\varepsilon)$ [cf.~Eq.~(\ref{PrbLES})] as a function of $t$ for the same
   values of $\delta\kappa$ as in panels (a)-(c).
   Other parameters: $N=1000$, $\kappa_i=0.4$ and
   $\delta\kappa_c=0.8$ [see Eq.~(\ref{Crtkp})].}
   \label{EvLSE}
 \end{figure*}

The agreement between the separatrix and the critical energy of the ESQPT implies that the logarithmic divergence in the DOS is a consequence of the saddle point in the underlying classical counterpart. 
To see this, we consider the semiclassical approximation of the DOS 
in the subspace with $N_{-1}=N_1$ \cite{Feldmann2021},
\be\label{DOSHc}
   \frac{\rho_{cl}(\varepsilon)}{N}=\frac{1}{(2\pi)^3}\int
   \mathcal{D}\bm{\xi}\delta(\zeta_{-1}-\zeta_1)\delta(\varepsilon-H_{cl}),
\ee
where $\mathcal{D}\bm{\xi}=\prod_md\zeta_md\varphi_m\delta(\sum_k\zeta_k-1)$.
Here, we would like to point out that $\rho_{cl}(\varepsilon)$ provides the smooth component 
in the Gutzwiller trace formula \cite{Gutzwiller2013}. 
The integral in the above equation can be carried out 
by employing the property of the delta-function and
the final results for different control parameters $\kappa$ are plotted as the red solid lines
in Figs.~\ref{SpDos}(b) and \ref{SpDos}(c). 
We see an excellent agreement between the numerical data and the behavior of 
$\rho_{cl}(\varepsilon)/N$.
In particular, $\rho_{cl}(\varepsilon)/N$ shows an obvious singularity at the critical 
energy of the ESQPT.

The existence of the saddle point in the classical system also explains the localization at the critical 
energy of the ESQPT.
To see this, we consider the localization behavior of an eigenstate $|\varepsilon_n\ra$ 
in the classical phase space, which can be visualized by the Husimi distribution \cite{Husimi1940}
\be
   Q(\zeta_0,\varphi)=|\la\varepsilon_n|\bm{\xi}\ra|^2.
\ee
It is known that the Husimi distribution $Q(\zeta_0,\varphi)$ offers the 
quasiprobability distribution of $|\varepsilon_n\ra$ in the 
classical phase space with canonical variables $(\zeta_0,\varphi)$.

Figs.~\ref{Hsprb}(b)-(d) plot the Husimi distribution for the eigenstates with energy below, at,
and above the crtical energy of the ESQPT.
For the eigenstate with the critical energy of the ESQPT [Fig.~\ref{Hsprb}(c)], 
the Husimi distribution displays a highly concentrated feature in the phase space.
By contrast, the Husimi distribution for the eigenstates with energy 
below and above the ESQPT is extended in the phase space, as seen in Figs.~\ref{Hsprb}(b) and
\ref{Hsprb}(d), respectively. 
The behaviors of the Husimi distribution for different eigenstates are consistent with 
the variation of the participation ratio with the eigenenergy.

\section{Loschmidt echo spectrum and ESQPT} \label{thirdS}

In the rest of the article, we analyze the dynamical signatures of the ESQPT by means of the 
Loschmidt echo spectrum.
To this end, we assume that the system is initially prepared in the ground state 
$\rho(0)=|\psi_0\ra\la\psi_0|$ of $\hat{H}_i$ with $\kappa=\kappa_i$.
Then, at $t=0$, we suddenly change the control parameter 
to a new value $\kappa=\kappa_i+\delta\kappa$ and focus on the dynamics of 
the system governed by the postquench Hamiltonian $\hat{H}_f$.
As the changing of the control parameter $\kappa$ leads to the varying energy in the system, 
one can drive the system crossing different phases of the ESQPT 
by tuning the value of $\delta\kappa$.
The critical quench strength, denoted by $\delta\kappa_c$, is defined as the 
one that takes the system to the critical energy of the ESQPT.
Employing the mean-field approximation, one can find $\delta\kappa_c$ is given by
\be\label{Crtkp}
  \delta\kappa_c=1-\frac{\kappa_i}{2},
\ee
where $0\leq\kappa_i\leq2$.
At this point,  it is worth pointing out that our conclusions of this work are independent of 
the choice of $\kappa_i$ as long as $0\leq\kappa_i\leq2$. 
Here, as we are focused only on investigating the dynamical features of ESQPT, the analysis on 
the cases with $\kappa_i>2(<0)$ are left for future study.
We stress that employing the Loschmidt echo spectrum to explore the quantum many-body dynamics 
for the critical line crossing quenches in the spinor BECs remains an open question. 
It was experimentally observed that the spin-1 BEC undergoes dynamical phase transitions as 
the control parameter quenches from the antiferromagnatic (AFM) region \cite{TianT2020}.
Hence, one can expect to find an abrupt change in the behavior of the Loschmidt echo spectrum
for quenches that cross the critical line.

After quench, the state of the system at time $t$ is $\rho(t)=e^{-i\hat{H}_ft}\rho(0)e^{i\hat{H}_ft}$.
To extract the dynamical signatures of the ESQPT from $\rho(t)$, we use the 
Loschmidt echo spectrum \cite{Wong2022} 
\be
  L_n(t)=\mathrm{Tr}[\rho_n\rho(t)]
       =\left|\sum_\alpha\la\psi_n|\alpha\ra\la\alpha|\psi_0\ra e^{-iE_\alpha t}\right|^2,
\ee
where $\rho_n=|\psi_n\ra\la\psi_n|$ is the $n$th eigenstate of $\hat{H}_i$ 
with associated energy $E_n$ 
and $|\alpha\ra$ is the $\alpha$th eigenstate of the postquench Hamiltonian $\hat{H}_f$ 
with the corresponding eigenenergy $E_\alpha$, so that $\hat{H}_f|\alpha\ra=E_\alpha|\alpha\ra$.

The Loschmidt echo spectrum measures how the time evolved state spreads 
in the eigenstates of the initial Hamiltonian and its ability to probe the dynamical 
quantum phase transitions in quantum many-body systems was investigated 
in a very recent work \cite{Wong2022}. 
Since the ESQPT involves the entire energy spectrum 
of the system, we would expect that more dynamical signatures 
of the ESQPT can be revealed by the Loschmidt echo spectrum.
Notice that the zero component of the Loschmidt echo spectrum, $L_0$,  is the 
well-known survival probability, which has been widely used 
as a dynamical detector of ESQPTs
\cite{Santos2015,Santos2016,Relano2008,Relano2009,PerezF2011,QianW2017,Kloc2018,
Stransky2021,Kloc2021}. 

Figs.~\ref{EvLSE}(a)-(c) demonstrate how the Loschmidt echo spectrum evolves as a function of $t$
for different quench strength $\delta\kappa$ with the system size $N=1000$ 
and $\kappa_i=0.4$, which determines $\delta\kappa_c=0.8$.
For the case with $\delta\kappa<\delta\kappa_c$, as illustrated in Fig.~\ref{EvLSE}(a), 
the small quench strength can only scatter the quenched state into the 
low excited states of the initial Hamiltonian. 
As a consequence, the Loschmidt echo spectrum is highly concentrated in the
low $H_i$ eigenenergies, irrespective of the time.
Moreover, small $\delta\kappa$ value also leads to the revival of the initial state, which results in
the regular oscillations in $t$ of the Loschmidt echo spectrum. 
With increasing $\delta\kappa$, more excited states are involved and the 
number of excited states contributing to the Loschmidt echo spectrum increases.
Once the critical quench $\delta\kappa_c=0.8$ is reached, the Loschmidt echo spectrum 
quickly spreads out over a larger number of excited eigenenergies and shows a complex dependence
on the time, as seen in Fig.~\ref{EvLSE}(b).
The fast spread of the Loschmidt echo spectrum at the critical quench can be considered as a result
of the saddle point in the corresponding classical system, which gives rise to 
the dynamical instability in the 
quantum system \cite{Silvia2018,Hummel2019,Cameo2020,XuTR2020,Hashimoto2020,Kidd2021}.
As $\delta\kappa$ increases further, such as $\delta\kappa=1.2$ case shown in 
Fig.~\ref{EvLSE}(c), the Loschmidt echo spectrum  also distributes  
over a wider range of excited eigenenergie, while it initially shows a regular oscillations which
is followed by an irregular pattern at later times.
The regular behavior in the evolved Loschmidt echo spectrum is a consequence of the periodic
trajectories in the underlying classical dynamics, while the long time irregular oscillations are due to
the quantum interference effect on the evolution of the Loschmidt echo spectrum.

 \subsection{Energy distribution of the evolved state}

The dynamical features observed in Figs.~\ref{EvLSE}(a)-(c) 
are more visible in Figs.~\ref{EvLSE}(d)-(f), where we plot the time evolution of 
the energy distribution of the evolved state weighted by components $L_n(t)$,
\be \label{PrbLES}
  P_t(\varepsilon)=\sum_nL_n(t)\delta(\varepsilon-\varepsilon_n).
\ee
It can be considered as the vertical cutting version of the evolved $L_n(t)$ at a fixed time and
satisfies $\int d\varepsilon P_t(\varepsilon)$=1.
We see that the evolution of $P_t(\varepsilon)$ shows an obvious difference between 
two phases of the ESQPT.
Hence, different phases of the ESQPT can be distinguished by the distinct behaviors in the evolved $P_t(\varepsilon)$ for $\delta\kappa<\delta\kappa_c$
and $\delta\kappa>\delta\kappa_c$, respectively. 
In addition, the singular behavior of $P_t(\varepsilon)$ at the critical quench stands as a
faithful figure of merit for diagnosing the presence of the ESQPT.

A main feature exhibited by $P_t(\varepsilon)$ is the different degrees of extension over the excited 
eigenenergies, which can be quantified by the variance of $P_t(\varepsilon)$ 
with the definition given by
\be
   \Sigma(t)=\int d\mathcal{\varepsilon}P_t(\varepsilon)[\varepsilon-\la\varepsilon\ra]^2,
\ee
where $\la\varepsilon\ra=\int d\varepsilon P_t(\varepsilon)\varepsilon$ 
is the averaging of $\varepsilon$.

  \begin{figure}
   \includegraphics[width=\columnwidth]{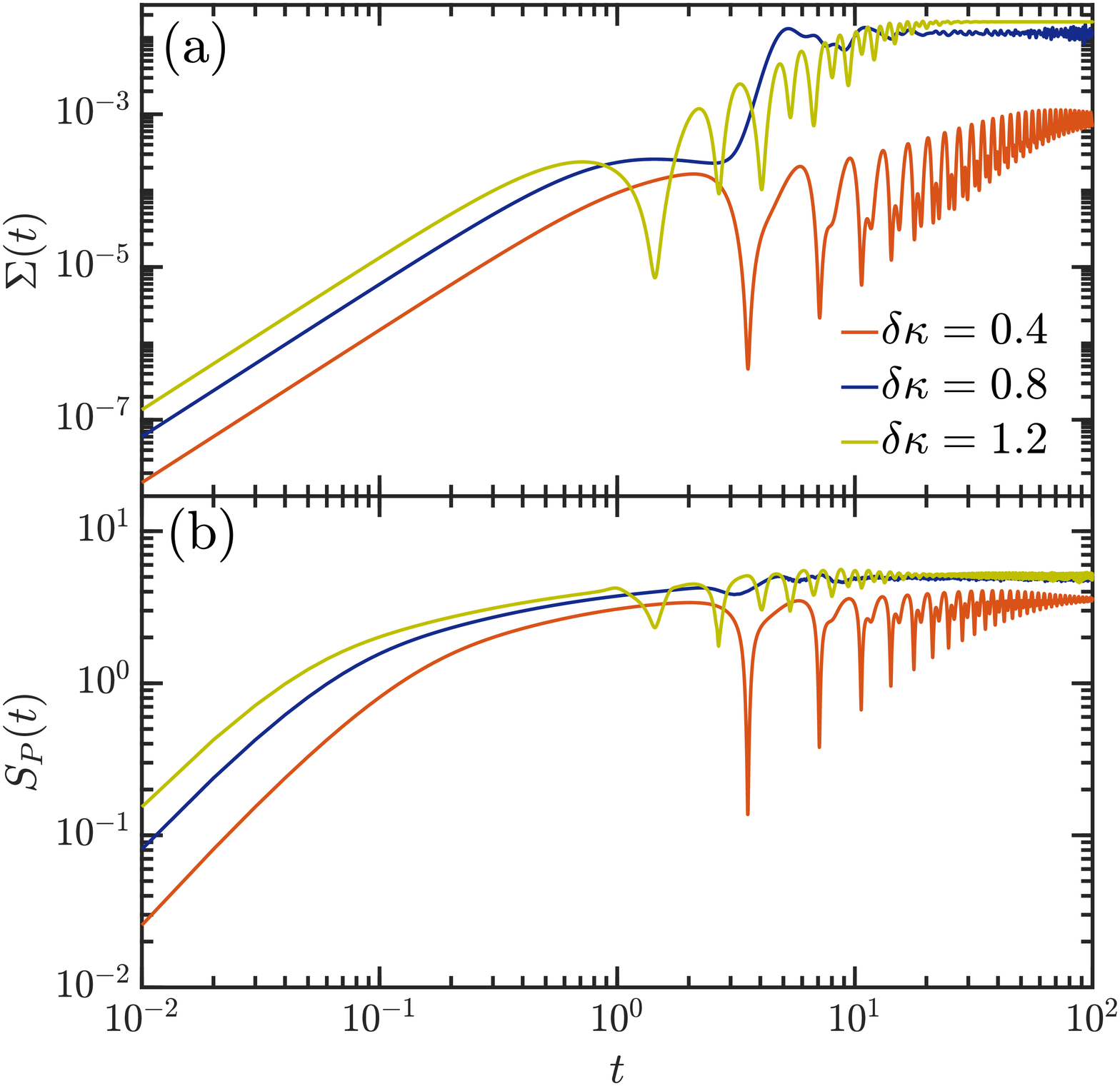}
   \caption{(a) $\Sigma(t)$ as a function of $t$ for several quench strengths $\delta\kappa$. 
   (b) Time evolution of the energy distribution entropy $S_P(t)$ for different values of $\delta\kappa$.
   Other parameters: $N=1000$, $\kappa_i=0.4$ and
   $\delta\kappa_c=0.8$ [see Eq.~(\ref{Crtkp})].}
   \label{SgmSp}
 \end{figure}

Fig.~\ref{SgmSp}(a) plots the variance of $\Sigma(t)$ with increasing $t$ for three different quench
strengths, which are below, at, and above the critical quench strength. 
Overall, the initial growth of $\Sigma(t)$ is followed by a small fluctuation around 
the saturation value, regardless of the strength of the quench.
However, the way of $\Sigma(t)$ growth depends strongly on the quench strength.
For $\delta\kappa$ below the critical value, $\Sigma(t)$ increases with larger oscillations 
and takes a long time to reach its saturation value with small oscillations.
On the other hand, even though the growth of $\Sigma(t)$ still shows many oscillations for 
$\delta\kappa>\delta\kappa_c$ case, it saturates to a larger saturation value at a short time scale.
At the critical quench $\delta\kappa=\delta\kappa_c$, $\Sigma(t)$ undergoes a fast growth which
then quickly saturates to its saturation value with tiny fluctuations. 
We further note that the saturation values of $\Sigma(t)$ for $\delta\kappa\geq\delta\kappa_c$
are almost independent of the quench strength.

The rapid increase in $\Sigma(t)$ at the critical quench can be explained as follows. 
The ESQPT associated with the saddle point in the underlying classical system, which leads to
the instability in the quantum many-body dynamics.
Consequently, the evolved state $\rho(t)$ is very spread out in the 
spectrum of the initial Hamiltonian in an extremely short time, 
as observed in Figs.~\ref{EvLSE}(b) and \ref{EvLSE}(e).
This results in the rapid growth in the variance of the energy distribution.

The observed features of $\Sigma(t)$ suggest that the underlying ESQPT leaves an imprint 
in the dynamics of quenched system.
Hence, the evolution of $\Sigma(t)$ can be employed to distinguish different phases of an ESQPT.
Moreover, the existence of the ESQPT can be efficiently signified by the rapid growth 
behavior in $\Sigma(t)$.

Figs.~\ref{EvLSE}(d)-(f) further demonstrate that the complexity of the evolved $P_t(\varepsilon)$
also shows a strong dependence on the quench strength.
To measure the complexity of $P_t(\varepsilon)$, we study the entropy of $P_t(\varepsilon)$:
\be  
   S_P(t)=-\int d\varepsilon P_t(\varepsilon)\ln P_t(\varepsilon).
\ee 
It varies in the interval $S_P(t)\in[0,\ln\Delta\varepsilon]$ with $\Delta\varepsilon$ 
is the width of the energy spectrum.
When the energy distribution is determined, that is, $P_t(\varepsilon)=1$, we have $S_P(t)=0$, while
$S_P(t)=\ln\Delta\varepsilon$ implies $P_t(\varepsilon)$ is uniform, namely 
$P_t(\varepsilon)=(\Delta\varepsilon)^{-1}$. 

Fig.~\ref{SgmSp}(b) shows the time evolution of $S_p(t)$ for the 
quench strengths that are the same as in Fig.~\ref{SgmSp}(a).
We see that the evolution of $S_P(t)$ is very similar to the behavior of $\Sigma(t)$, apart from 
small fluctuations around the saturation value.
We also note that $S_P(t)$ saturates at a very short time scale in comparison with $\Sigma(t)$.
However, due to the logarithmic definition in $S_P(t)$, 
the growth of $S_P(t)$ at the critical quench strength is not as rapid as in the case of $\Sigma(t)$.  
Nevertheless, the similarity between $S_P(t)$ and $\Sigma(t)$ means that both of them can be
used to identify the occurrence and different phases of the ESQPT.

  \begin{figure*}
   \includegraphics[width=\textwidth]{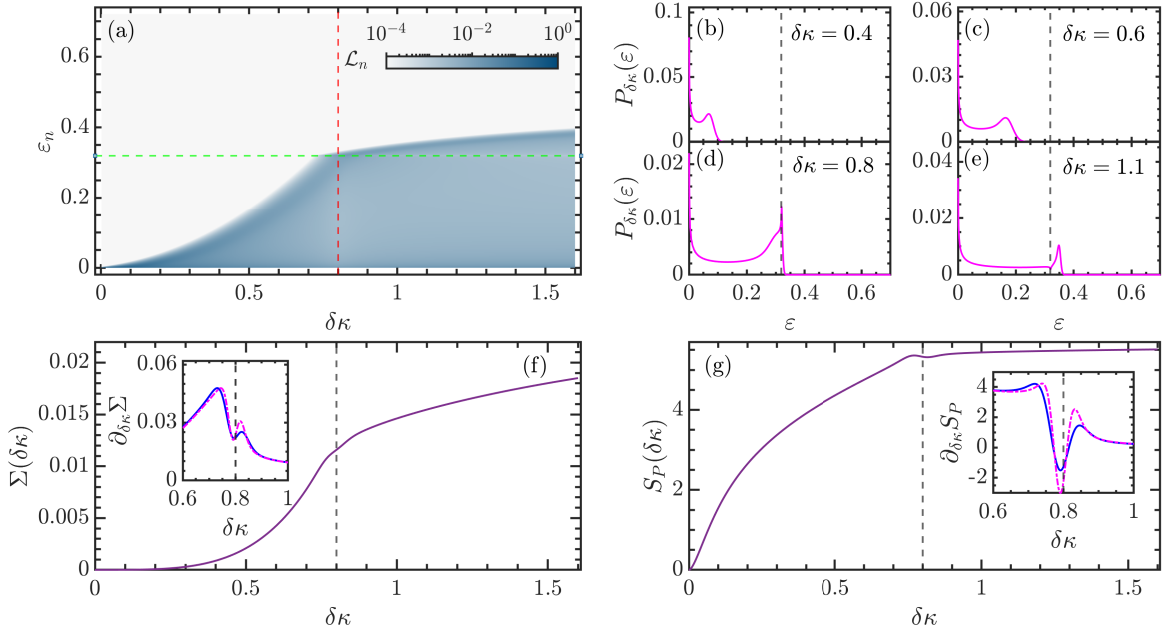}
   \caption{(a): Heat map of the long time averaged Loschmidt echo spectrum $\mathcal{L}_n$
   as a function of $\delta\kappa$ and $\varepsilon_n$ with system size $N=1000$.
   The horizontal dashed line marks the critical energy $\varepsilon_c$ of the ESQPT, while
   the vertical red dashed line is the critical quench strength from Eq.~(\ref{Crtkp}).
   (b)-(e): Energy distribution of the long time averaged state $P_{\delta\kappa}(\varepsilon)$
   for different values of $\delta\kappa$ with $N=1000$.
   The vertical dashed line in each panel denotes the critical energy $\varepsilon_c$ of the ESQPT,
   see Eq.~(\ref{CrtE}).
   (f): Variance of the energy distribution of the long time averaged state 
   as a function of $\delta\kappa$ for $N=1000$.
   The inset shows $\partial_{\delta\kappa}\Sigma$ as a function of $\delta\kappa$ for 
   $N=1000$ (blue solid curve) and $N=2000$ (magenta dot-dashed curve). 
   (g): Entropy of the energy distribution of the long time averaged state as a function of $\delta\kappa$
   for $N=1000$.
   The inset plots $\partial_{\delta\kappa}S_P$ as a function of $\delta\kappa$ for 
   $N=1000$ (blue solid curve) and $N=2000$ (magenta dot-dashed curve).
   The vertical dashed line in panels (f) and (g) is the critical quench strength $\delta\kappa_c$.
   Other parameters: $\kappa_i=0.4$ and $\delta\kappa_c=0.8$ obtained from Eq.~(\ref{Crtkp}).}
   \label{TavLES}
 \end{figure*}

\subsection{Long time averaged Loschmidt echo spectrum}

The signatures of the ESQPT are also reflected in the long time averaged Loschmidt echo spectrum,
which is calculated as
\begin{widetext}
\be \label{LTAV}
   \mathcal{L}_n=\lim_{T\to\infty}\frac{1}{T}\int_0^TL_n(t)dt
   =\lim_{T\to\infty}\frac{1}{T}\int_0^Tdt\sum_{\alpha,\beta}\la\psi_n|\alpha\ra\la\alpha|\psi_0\ra
     \la\beta|\psi_n\ra\la\psi_0|\beta\ra e^{-it(E_\alpha-E_\beta)}
    =\sum_\alpha|\la\psi_n|\alpha\ra|^2|\la\alpha|\psi_0\ra|^2,
\ee
\end{widetext}
where $H_f|\alpha\ra=E_\alpha|\alpha\ra$, $H_f|\beta\ra=E_\beta|\beta\ra$, and
we carried out the integration as the energy spectrum of $H_f$ is non-degenerate.
The $\mathcal{L}_n$ can be recognized as the overlap between the long time averaged state  
$\bar{\rho}=\lim_{T\to\infty}\int_0^T\rho(t)dt$ and 
the $n$th eigenstate $\rho_n=|\psi_n\ra\la\psi_n|$ of $H_i$, so that
$\mathcal{L}_n=\mathrm{Tr}(\bar{\rho}\rho_n)$. 
Here, we should point out that the final simplified form in the above equation 
only holds for the finite system. 
In the thermodynamic limit $N\to\infty$, as energy levels are degenerated 
at the ESQPT critical energy, the last equation in (\ref{LTAV}) is invalid.
It is also worth mentioning that the reciprocal $\mathcal{L}_0$ is the 
participation ratio of the initial state $|\psi_0\ra$ with respect to the eigenstates of $H_f$.

In Fig.~\ref{TavLES} (a), we plot the long time averaged Loschmidt echo spectrum 
$\mathcal{L}_n$ as a function of $\delta\kappa$ and $\varepsilon_n$ for 
$\kappa_i=0.4$ and $N=1000$.
We see that the ESQPT at $\delta\kappa_c=0.8$ leads to a remarkable change in the behavior of 
$\mathcal{L}_n$, which, in turn, can be used to diagnose the onset of the ESQPT. 
Moreover, $\mathcal{L}_n$ for $\delta\kappa>\delta\kappa_c$ also shows a dip at 
the ESQPT critical energy $\varepsilon_c$, as marked by the horizontal dashed line in the figure.
The dip observed in $\mathcal{L}_n$ at the critical energy for $\delta\kappa>\delta\kappa_c$ case 
is a consequence of the localization 
of the critical eigenstate $|\psi_c\rangle$ over the spectrum of $H_f$, which results in the
significant contribution to $\mathcal{L}_n$ 
are the states with small values of $|\langle\psi_c|\alpha\rangle|^2$.

More information about the effect of the ESQPT
are provided by the energy distribution of the long time averaged state weighted by $\mathcal{L}_n$,
\be
   P_{\delta\kappa}(\varepsilon)=\sum_n\mathcal{L}_n\delta(\varepsilon-\varepsilon_n).
\ee
Figs.~\ref{TavLES}(b)-(e) demonstrate the energy distribution $P_{\delta\kappa}(\varepsilon)$ for
several $\delta\kappa$ values.
One can clearly see that $P_{\delta\kappa}(\varepsilon)$ exhibits distinct behaviors in different phases
of the ESQPT.
In particular, due to the localization of the critical eigenstate $|\psi_c\ra$,
the critical value of $\mathcal{L}_n$ is dominated 
by the large $|\la\psi_c|\alpha\ra|^2$ terms, when taking the system to the ESQPT critical energy.
This means that, for the critical quench strength, the energy distribution 
has a sharp peak at the ESQPT critical energy, as observed in Fig.~\ref{TavLES}(d). 
This peak signals the ESQPT singularity and can be employed to probe
the occurrence of the ESQPT.

To quantitatively investigate the features of $P_{\delta\kappa}(\varepsilon)$, 
as we did for the time evolved energy distribution $P_t(\varepsilon)$, we study
its variance and entropy, which are, respectively, defined by
\begin{align}
 &\Sigma(\delta\kappa)=\int d\varepsilon P_{\delta\kappa}(\varepsilon)
[\varepsilon-\la\varepsilon\ra]^2,\notag  \\ 
&S_P(\delta\kappa)=-\int d\varepsilon P_{\delta\kappa}(\varepsilon)
\ln P_{\delta\kappa}(\varepsilon),
\end{align}
where $\la\varepsilon\ra=\int d\varepsilon P_{\delta\kappa}(\varepsilon)\varepsilon$.

In Fig.~\ref{TavLES}(f), we display $\Sigma(\delta\kappa)$ as a function of $\delta\kappa$ 
for $\kappa_i=0.4$ and $N=1000$.
The variance $\Sigma(\delta\kappa)$ increases with increasing $\delta\kappa$, but
it grows in different rate between two phases of the ESQPT.
The changing of the growth rate in $\Sigma(\delta\kappa)$ occurs 
at the ESQPT critical quench strength, due to 
the localized eigenstates around the ESQPT critical energy, which lead to
the energy distribution remains almost unchanged for the quenches around $\delta\kappa_c$.
This is confirmed by the derivative of the variance with respect to $\delta\kappa$, denoted by
$\partial_{\delta\kappa}\Sigma(\delta\kappa)$,
which is plotted in the inset of Fig.~\ref{TavLES}(f) for the cases of different system sizes.
As we can see, $\partial_{\delta\kappa}\Sigma(\delta\kappa)$ shows an abrupt decrease and
reaches its local minimal value near the critical quench strength.
Hence, the variance $\Sigma(\delta\kappa)$ succinctly captures the signatures of the ESQPT.

The dependence of entropy $S_P(\delta\kappa)$ on the quench strength 
is plotted in Fig.~\ref{TavLES}(g).
For $\delta\kappa<\delta\kappa_c$, $S_P(\delta\kappa)$ grows as $\delta\kappa$ increases, while
it saturates to a stationary value as $\delta\kappa$ increases for $\delta\kappa>\delta\kappa_c$.
This difference in the behavior of $S_P(\delta\kappa)$ allows us to use it to identify 
different phases of the ESQPT.
In the neighborhood of the critical quench strength $\delta\kappa_c$, we see a decrease
in $S_P(\delta\kappa)$.
This means that the entropy $S_P(\delta\kappa)$ can be used as a witness of the ESQPT.

The eigenstates localization associated with the ESQPT critical point 
is also the source of the decrease of the entropy near $\delta\kappa_c$.
The localized eigenstates around the ESQPT critical energy give rise to small 
values of $P_{\delta\kappa}(\varepsilon)$ for the critical quench, as seen 
in Fig.~\ref{TavLES}(d). 
Hence, the entropy $S_P(\delta\kappa)$ would have a local minimal value at $\delta\kappa_c$.
This is more visible in the inset of Fig.~\ref{TavLES}(f), where we show how the derivative of entropy
$\partial_{\delta\kappa}S_P$ evolves as a function of $\delta\kappa$ for different system sizes.
An obvious dip in the value of $\partial_{\delta\kappa}S_P$ appears near the critical quench strength,
confirming that the entropy $S_P(\delta\kappa)$ undergoes a decrease as the strength of quench passes
through its critical value. 
Increasing the system size $N$ tends to sharpen the dip 
and moving its location to the critical value of $\delta\kappa$. 

  \begin{figure*}
   \includegraphics[width=\textwidth]{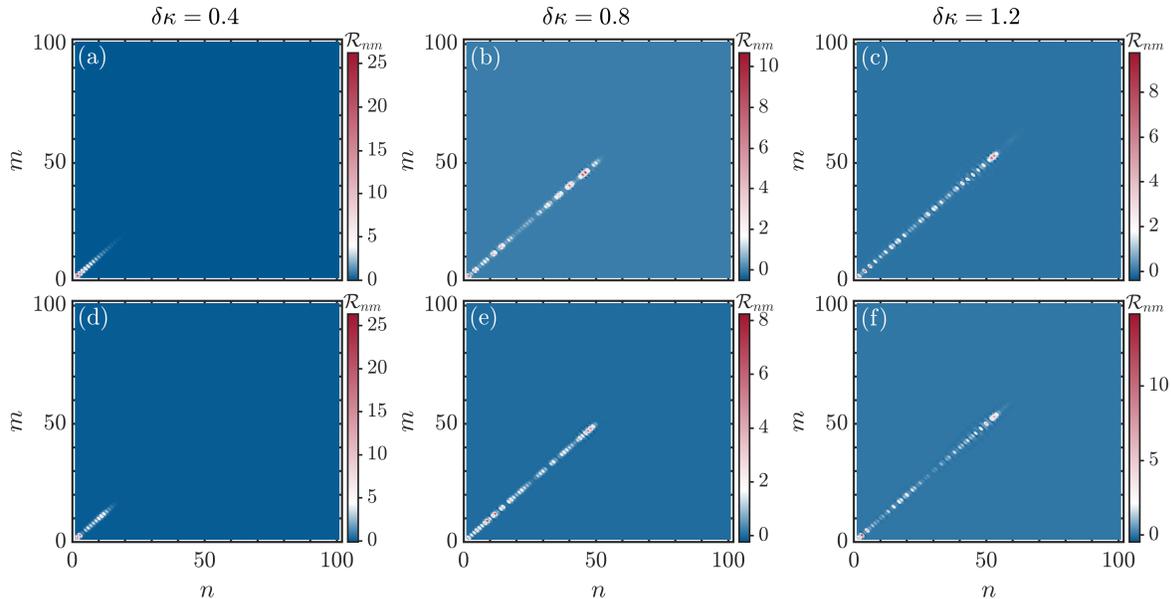}
   \caption{$\mathfrak{R}_{nm}$ in Eq.~(\ref{RealP}) with respect
   to the $n$th and $m$th eigenstates of the initial Hamiltonian 
   for three typical quenches at two times:
   $t=10$ (a)-(c) and $t=100$ (d)-(f).
   Other parameters: $N=200$, $\kappa_i=0.4$, 
   and $\delta\kappa_c=0.8$ obtained from Eq.~(\ref{Crtkp}).}
   \label{Tvft}
 \end{figure*}

It is also interesting to investigate whether the Loschmidt echo spectrum can provide
insights into the dynamics of $f_0=\la\hat{N}_0\ra/N$, which has been used to
study both ground state QPT \cite{Dag2018,XueM2018} 
and dynamical phase transition \cite{TianT2020} in spin-1 BEC.
The time-evolving $f_0$ is given by
\begin{widetext}
\begin{align}\label{DsyN}
  \la f_0(t)\ra=\frac{1}{N}\la\psi_0|e^{iH_ft}\hat{N}_0e^{-iH_ft}|\psi_0\ra
     =\frac{1}{N}\sum_{n,m}\mathcal{A}_n^\ast(t)\mathcal{A}_m(t)N_{0,nm}
     =\frac{1}{N}\left[\sum_n L_n(t)N_{0,nn}+2\sum_{n<m}\mathfrak{R}_{nm}\right],
\end{align}
\end{widetext}
where $N_{0,k\ell}=\la\psi_k|\hat{N}_0|\psi_\ell\ra$, 
$\mathcal{A}_{k}(t)=\la\psi_k|e^{-iH_ft}|\psi_0\ra$ is the amplitude of $L_k(t)$,
and
\be \label{RealP}
  \mathfrak{R}_{nm}=\mathrm{Re}
     \left[\mathcal{A}_{n}^\ast(t)\mathcal{A}_m(t)N_{0,nm}\right],
\ee 
denotes the real part of $\mathcal{A}_{n}^\ast(t)\mathcal{A}_m(t)N_{0,nm}$.
Fig.~\ref{Tvft} plots $\mathfrak{R}_{nm}$ for 
different quenches at different times.
We see that, regardless of the quench strength and time, the diagonal terms are
much larger compared to the off-diagonal terms.
This implies that the summation in Eq.~(\ref{DsyN}) is governed by the diagonal terms,
which are directly connected to the Loschmidt echo spectrum. 
The Loschmidt echo spectrum can therefore help us to understand the 
features that are exhibited in $f_0$ dynamics.

As a final remark of this section, we would like to point out that 
the Loschmidt echo spectrum can be experimentally detected
by recent proposed protocols \cite{Swingle2016,Vasilyev2020,DagS2022}. 
Moreover, we checked that 
the dynamical signatures of the ESQPT revealed in time evolved and long time averaged
Loschmidt echo spectrum can be observed in a several hundred milliseconds for the typical spin-spin
interaction strength $|c|/h=2\pi\times 9\ \mathrm{Hz}$ \cite{SKurn2013}.
These facts indicate that the experimental verification of our findings in the 
spin-1 BEC platform is achievable.

 \section{Conclusions} \label{summary}
 
 In summary, by investigating the dynamical signatures of the ESQPT in the spin-$1$ BEC via the
 Loschmidt echo spectrum, we showed the usefulness of the Loschmidt echo spectrum 
 in the studying of ESQPTs in quantum many-body systems.
 We explored how the ESQPT gets reflected in the
 behaviors of the Loschmidt echo spectrum and associated energy distribution.
   
 The ESQPT in the spin-$1$ BEC is characterized by the logarithmic divergence of the density of states,
 which results in the localization of the eigenstates around the ESQPT. 
 A detailed study of the classical limit of the system reveals that the presence of the ESQPT 
 and the corresponding characterizations are a consequence of the saddle point in the classical system.

 The analysis of the time evolution of the Loschmidt echo spectrum unveils that 
 the underlying ESQPT leads to a drastic change in the behavior of the Loschmidt echo spectrum.
 In particular, we have seen that the singular behavior in the evolution of the 
 Loschmidt echo spectrum signals the onset of the ESQPT.
 We further showed that the occurrence and different phases of the ESQPT 
 were clearly identified in the properties of 
 the energy distribution of the evolved state weighted by the components 
 of the Loschmidt echo spectrum.
 In addition, an explicit examination of the long time averaged Loschmidt echo 
 and associated energy distribution demonstrates that they
 can also be utilized to probe the ESQPT.
 
 Although our conclusions are verified for spin-$1$ BEC, we would expect that qualitatively 
 similar results should also be found in other systems that exhibit 
 the logarithmic divergence in their density of states. 
 A natural extension of the present work is to study the classification of the ESQPTs 
 by performing scaling analyses on the critical behaviors of the Loschmidt echo spectrum.
 Classifying the ESQPTs is non-trivial and remains an open question.
 It would be also interesting to explore the relationship between 
 the Loschmidt echo spectrum and the
 ESQPTs, which are signified by the discontinuity in the 
 derivative of the density of states \cite{Cejnar2021,Stransky2014,Stransky2016}. 
 Finally, our findings extend the usefulness of the Loschmidt echo spectrum in studying of the
 dynamical quantum phase transitions \cite{Wong2022} to ESQPTs, and open
 a different way to obtain a better understanding on the dynamical features of ESQPTs.

 \acknowledgements
 
 This work is supported by the 
 Zhejiang Provincial Nature Science Foundation (Grants No.~LY20A050001 and No.~LQ22A040006);
 the National Science Foundation of China under Grant No.~11805165;
 and the Slovenian Research Agency (ARRS) under Grant No.~J1-4387.

\bibliographystyle{apsrev4-1}
\bibliography{SPBEC}

\end{document}